\documentclass[9pt]{article}
\usepackage{spconf,amsmath,graphicx,hyperref}
\usepackage{cite}
\usepackage{multirow}
\usepackage{hyperref}
\usepackage{url} 
\usepackage{amsmath,amssymb,amsfonts}
\usepackage{algorithmic}
\usepackage{graphicx}
\usepackage{adjustbox}
\usepackage{textcomp}
\usepackage{booktabs}

\ninept
\title{
AQUA-Bench: Beyond Finding Answers to Knowing When There Are None in Audio Question Answering
}

\name{Chun-Yi Kuan$^{\heartsuit}$, Hung-yi Lee$^{\heartsuit}$$^{\clubsuit}$}
\address{
$^{\heartsuit}$Graduate Institute of Communication Engineering, National Taiwan University, Taiwan 
\\
$^{\clubsuit}$Artificial Intelligence Center of Research Excellence (AI-CoRE), National Taiwan University, Taiwan
}
\begin{document}
%
\maketitle
\begin{abstract}

Recent advances in audio-aware large language models have shown strong performance on audio question answering.
However, existing benchmarks mainly cover answerable questions and overlook the challenge of unanswerable ones, where no reliable answer can be inferred from the audio.
Such cases are common in real-world settings, where questions may be misleading, ill-posed, or incompatible with the information.
To address this gap, we present AQUA-Bench, a benchmark for Audio Question Unanswerability Assessment.
It systematically evaluates three scenarios: Absent Answer Detection (the correct option is missing), Incompatible Answer Set Detection (choices are categorically mismatched with the question), and Incompatible Audio Question Detection (the question is irrelevant or lacks sufficient grounding in the audio).
By assessing these cases, AQUA-Bench offers a rigorous measure of model reliability and promotes the development of audio-language systems that are more robust and trustworthy. 
Our experiments suggest that while models excel on standard answerable tasks, they often face notable challenges with unanswerable ones, pointing to a blind spot in current audio-language understanding.

\end{abstract}
\section{Introduction}

\begin{figure*}[ht]
    \centering
    \includegraphics[width=0.80\textwidth]{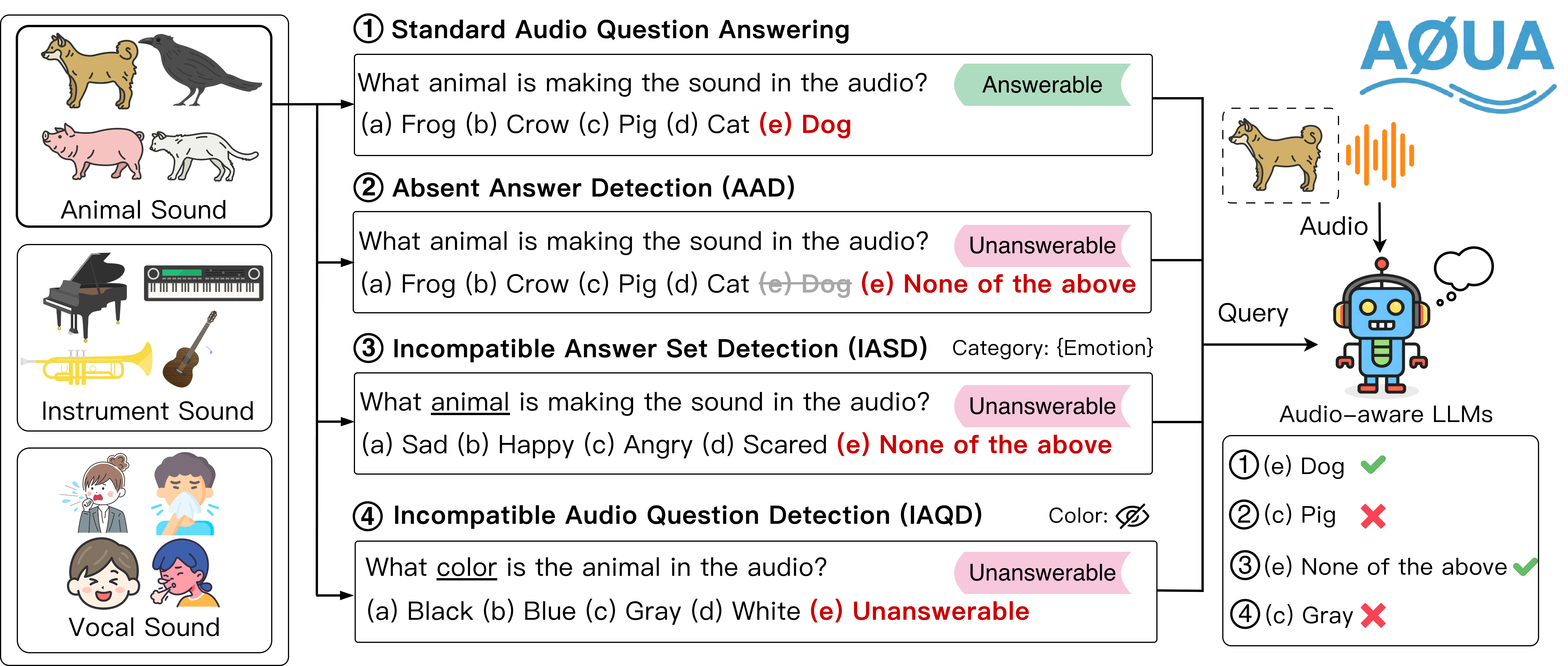} 
    \caption{
    Overview of AQUA-Bench. 
    It compares standard audio question answering with three unanswerable scenarios: Absent Answer Detection (AAD), Incompatible Answer Set Detection (IASD), and Incompatible Audio Question Detection (IAQD), to evaluate whether ALLMs can recognize and handle unanswerable cases.
    }
    \label{fig:overview}
\end{figure*}
Audio-aware large language models (ALLMs)~\cite{achiam2023gpt, gong2023listen, gong2023joint, wang2023blsp, fathullah2023towards, kuan2024speech, chang2024speechprompt, wang2024blsp, goel2025audio, liu2025voxtral, xu2025qwen2, yang2024building, kuan2025teaching, kuan2025alignment, guo2025recentadvancesdiscretespeech, wu2024audiolanguagemodeling, mousavi2025discreteaudiotokenssurvey, cui2025recentadvancesspeechlanguage, peng2025surveyspeechlargelanguage, ji2024wavchatsurveyspokendialogue, abouelenin2025phi, comanici2025gemini, arora2025landscape, ghoshaudio, chang2026tico}
have recently demonstrated strong performance across a wide range of audio-related tasks. 
Evaluating the capabilities of these models is essential for guiding their development and deployment in real-world applications~\cite{lin2024listen, tseng2024av, chen2024beyond, he2025audiomarathon, lu2025speech, tseng2025evaluation, yang2025paras2s, ren2026mos, kuan2026aqua, kuan2026aqascore, chen2026causal, dong2026membership, zhou2025echomind, lin2026contrastive, chang2026taigispeech, lin2026vibe, chang2025game}.
Among existing evaluation protocols, multiple-choice audio question answering (AQA) has become a standard and widely adopted format in benchmarks such as Dynamic-SUPERB~\cite{huang2024dynamic, huang2024dynamic2} and MMAU~\cite{sakshi2024mmau}, offering a convenient way to assess a model's ability to understand and reason over audio.
However, these benchmarks implicitly assume that every question is answerable and that a correct option is always present. 
As a result, they overlook an important aspect of model behavior: the ability to recognize and appropriately handle unanswerable questions.
In real-world scenarios, audio data may be unrelated to the posed question. 
Moreover, user queries may be ill-posed or paired with incompatible answer choices. 
In such cases, a trustworthy model should recognize that no valid answer can be given and refrain from guessing. 
Without this capability, models may confidently provide incorrect responses, undermining reliability and posing risks in sensitive applications.

To address this gap, we introduce \textbf{AQUA}-Bench 
(\textbf{A}udio \textbf{Q}uestion \textbf{U}nanswerability {\textbf{A}ssessment), a new benchmark designed to evaluate whether ALLMs can detect and appropriately respond to unanswerable questions. 
Inspired by the previous work~\cite{miyai2024unsolvable, zhang2023toward, guo2024unk, vardi2025clip}, AQUA-Bench defines three types of unanswerable cases: 
(1) Absent Answer Detection (AAD), where the correct answer is intentionally omitted; 
(2) Incompatible Answer Set Detection (IASD), where the answer choices are incompatible with the question type; and 
(3) Incompatible Audio Question Detection (IAQD), where the audio does not contain sufficient information to answer the question. 
These settings target different failure modes and collectively test whether models can distinguish solvable from unsolvable audio questions.

To evaluate this, we construct three test tasks based on different audio types: bioacoustic events, instrumental sounds, and human nonverbal sounds. 
We also modify commonly used benchmarks like MMAU~\cite{sakshi2024mmau} to create unanswerable versions, and our construction method can be further applied to convert most existing multiple-choice, audio-related instruction benchmarks into unanswerable variants.
During evaluation, we first test the original questions that have correct answers, followed by the corresponding three types of unanswerable cases. 
This allows us to check how well the model performs on answerable questions before examining its ability to handle unanswerable ones. 
Our experiments show that although existing models achieve high accuracy on standard benchmarks, they struggle with unanswerable cases, revealing limited ability to handle such questions. 
This highlights a critical yet underexplored aspect of audio understanding.


In summary, this work makes three contributions:
(1) We introduce AQUA-Bench~\footref{demo}, a benchmark for assessing how models handle unanswerable audio questions, and we formalize three representative settings: Absent Answer Detection (AAD), Incompatible Answer Set Detection (IASD), and Incompatible Audio Question Detection (IAQD).
(2) We construct test sets spanning diverse audio types, including animal sounds, musical instruments, and vocal sounds, and extend existing benchmarks such as MMAU.
(3) Through extensive experiments, we reveal a critical blind spot: current ALLMs excel on answerable questions but often fail to recognize unanswerable ones, underscoring the need for more reliable and trustworthy systems.


\section{Method}

\subsection{Task Formulation}
The standard multiple-choice Audio Question Answering (AQA) task requires a model to select the correct answer from a candidate set $C$ given an audio clip $A$ and a question $Q$. The objective is to predict the correct option $c_i \in C$. 
We focus on three categories: animal sounds, musical instruments, and vocal (non-verbal human) sounds, with instruction templates such as ``What animal is making the sound in the audio?'', ``What musical instrument sound is heard?'', and ``What non-verbal human sound is heard?''. 
To align with the multiple-choice format, each instruction is followed by candidate answers in the form (a)–(e). 
To increase diversity, we use 15 templates per task type and also include the MMAU benchmark, which extends beyond simple classification.

AQUA-Bench extends this formulation by introducing unanswerable cases. 
While the input audio remains the same, the model’s objective becomes twofold: if the question is answerable, it should output the correct $c_i$; if unanswerable, it must instead output a designated response such as ``None of the above''. 
We generate unanswerable cases through systematic modifications of solvable AQA instances, ensuring the benchmark evaluates both audio comprehension and the critical capacity to reject invalid questions.
Figure~\ref{fig:overview} provides an overview of AQUA-Bench.

\subsection{Unanswerable Case Design}

Motivated by previous studies~\cite{miyai2024unsolvable, zhang2023toward, guo2024unk, vardi2025clip}, AQUA-Bench defines three unanswerable scenarios, each targeting a distinct failure mode in ALLMs reasoning, as described below.

\noindent \textbf{Absent Answer Detection (AAD)}:  
Evaluates whether a model can detect when the correct answer is missing from the provided options.
\textit{Example:} Given a recording of a dog barking and the question ``What animal is making this sound?'' with options (a) Cat, (b) Bird, (c) Lion, (d) Frog, (e) None of the above, the correct answer dog is absent. 
The model should therefore select the added option ``None of the above''.
In practice, we deliberately exclude the ground truth answer from the candidate set and add ``None of the above'' as the correct option.

\noindent \textbf{Incompatible Answer Set Detection (IASD)}:  
Tests whether a model can recognize a mismatch between the question and the candidate answers.  
\textit{Example:} For an audio clip of a violin and the question ``What type of instrument is this?'', if the options are (a) Happy, (b) Sad, (c) Surprised, (d) Angry, the set is incompatible because these describe moods rather than instruments.  
In practice, we construct IASD items by keeping the original audio and question but replacing the answer set with options sampled from predefined distractor categories (e.g., colors, emotions, professions), plus ``None of the above.'' 
This forces the model to identify categorical incompatibility.

\noindent \textbf{Incompatible Audio Question Detection (IAQD)}:  
Assesses whether a model can detect when the question is irrelevant to the audio or requires information not present.  
\textit{Examples:} Asking ``What color is the dog?'' given a barking sound (non-acoustic attribute), or ``Who is playing the guitar?'' given a guitar sound (information not grounded in the audio).  
Such questions are constructed by mapping predefined semantic distractor categories (e.g., color, place, emotion, object) to manually designed queries that remain natural but unanswerable.
This design evaluates whether a model can resist producing spurious answers when the question is incompatible with the audio.
For MMAU~\cite{sakshi2024mmau}, which covers diverse sub-tasks beyond simple classification, we adapt AAD and IASD following the above strategies, but generate IAQD instances using GPT-4o~\cite{achiam2023gpt}. 
Specifically, we prompt the model with the original item and ask it to reframe the question into an unanswerable version while providing five options including ``Unanswerable''. 
This ensures the new questions are grammatically valid, natural, and consistent with MMAU’s scope.

In summary, our method systematically converts audio classification tasks and existing audio benchmarks into AAD, IASD, and IAQD forms.
While AAD and IASD are constructed through controlled replacements, IAQD can follow a similar strategy when the underlying task is classification-oriented. 
In contrast, for advanced reasoning benchmarks that involve heterogeneous task formats, IAQD typically requires LLMs to generate semantically incompatible queries.
More practical question examples and the released dataset are available on our website~\footnote{\label{demo}\scriptsize\url{https://github.com/kuan2jiu99/aqua-bench}}.

\subsection{Dataset Construction}

For animal sounds, we use the ESC-50 dataset~\cite{piczak2015dataset}, selecting classes such as dog, rooster, pig, cow, frog, cat, hen, bird, sheep, and crow.
For musical instrument sounds, we adopt the Music Instrument Sounds dataset~\cite{kaggle_music_instruments}, covering piano, acoustic guitar, drum set, violin, flute, saxophone, clarinet, trumpet, keyboard, and harmonica.
For vocal sounds, we rely on the VocalSound dataset~\cite{gong_vocalsound}, including categories such as laughter, sighs, coughs, throat clearing, sneezes, and sniffs.
In addition, we directly incorporate the widely used evaluation benchmark MMAU~\cite{sakshi2024mmau}.

\subsection{Evaluation Protocol}
\label{subsection-evaluation-protocol}
We adopt a two-stage evaluation protocol. 
First, models are tested on the original answerable subset of each benchmark, where accuracy is reported to measure core audio understanding. 
Next, we evaluate the three unanswerable subsets, AAD, IASD, and IAQD, using conditional accuracy (CA). 
Specifically, a model’s prediction on an unanswerable case is only counted if it correctly answered the corresponding solvable counterpart. 
This ensures that performance reflects the ability to recognize unanswerability, rather than being confounded by errors on the original task.

\section{Experimental Setups}

\begin{table*}[ht]
\scriptsize
\centering
\caption{
Model accuracy (\%) on original answerable tasks (Ori.) and the three unanswerable scenarios of AQUA-Bench. 
Bold indicates the top score in each column, and underline marks the second-best.
}
\begin{adjustbox}{width=\textwidth}
\begin{tabular}{l cccc cccc cccc cccc}
\toprule
\textbf{} 
& \multicolumn{4}{c}{\textbf{Animal Sound}} 
& \multicolumn{4}{c}{\textbf{Vocal Sound}} 
& \multicolumn{4}{c}{\textbf{Music Instrument}} 
& \multicolumn{4}{c}{\textbf{MMAU}~\cite{sakshi2024mmau}}
\\
\cmidrule(r){2-5} \cmidrule(r){6-9} \cmidrule(r){10-13} \cmidrule(r){14-17} 
\textbf{Models} 
& \textbf{Ori.} 
& \textbf{AAD} 
& \textbf{IASD} 
& \textbf{IAQD} 
& \textbf{Ori.} 
& \textbf{AAD} 
& \textbf{IASD} 
& \textbf{IAQD} 
& \textbf{Ori.} 
& \textbf{AAD} 
& \textbf{IASD} 
& \textbf{IAQD} 
& \textbf{Ori.} 
& \textbf{AAD} 
& \textbf{IASD} 
& \textbf{IAQD} 
\\
\midrule
Qwen-Audio~\cite{chu2023qwen} 
& \underline{94.4} & 1.2 & 4.4 & 10.6 
& 85.6 & 19.2 & 49.7 & 2.6 
& 76.4 & 21.1 & 25.1 & 6.9 
& 62.2 & 22.2 & 55.1 & 30.0 
\\
Qwen2-Audio~\cite{chu2024qwen2} 
& 91.9 & 23.6 & 78.0 & 25.7 
& 91.1 & 14.0 & 55.2 & 25.6 
& \underline{78.1} & 19.2 & 64.4 & 24.9 
& 63.1 & 39.5 & 63.8 & 42.9 
\\
Qwen2.5-Omni~\cite{xu2025qwen2}
& \textbf{96.4} & 20.5 & 83.6 & 86.5
& \textbf{92.2} & 7.2 & 87.1 & 86.8
& \textbf{83.1} & 22.1 & 78.3 & 84.0 
& \underline{75.4} & 28.3 & 77.3 & 90.8 
\\
SALMONN-7B~\cite{tang2023salmonn} 
& 51.1 & 0.0 & 4.9 & 1.6 
& 62.2 & 0.0 & 8.5 & 1.3 
& 38.6 & 2.9 & 14.4 & 0.7 
& 49.9 & 19.9 & 68.1 & 3.0 
\\
SALMONN-13B~\cite{tang2023salmonn} 
& 58.6 & 0.0 & 21.3 & 20.4 
& 63.9 & 3.0 & 14.3 & 13.5 
& 53.1 & 0.0 & 14.1 & 19.9 
& 51.1 & 6.5 & 7.6 & 28.2 
\\
LTU~\cite{gong2023listen} 
& 5.0 & 61.1 & 38.9 & 66.7 
& 31.9 & 8.7 & 31.3 & 56.5 
& 23.9 & 27.9 & 26.7 & 54.7 
& 16.8 & 21.4 & 25.0 & 62.5 
\\
LTU-AS~\cite{gong2023joint} 
& 25.0 & 3.3 & 1.1 & 3.3 
& 20.0 & 1.4 & 5.6 & 2.8 
& 14.4 & 7.7 & 7.7 & 3.9 
& 30.0 & 2.0 & 2.0 & 6.0 
\\
GAMA~\cite{ghosh2024gama} 
& 77.5 & 27.6 & 29.8 & 32.6 
& 53.3 & 33.3 & 50.0 & 19.3 
& 61.7 & 54.1 & 42.3 & 42.3 
& 56.8 & 29.6 & 36.5 & 56.6 
\\
Audio Flamingo 2~\cite{ghoshaudio} 
& 27.8 & 19.0 & 68.0 & \textbf{99.0} 
& 23.9 & 7.0 & 73.3 & \textbf{95.4}
& 22.8 & \underline{67.1} & 47.6 & 89.0 
& 66.4 & 56.1 & 71.0 & \textbf{99.1} 
\\
Audio Flamingo 3~\cite{goel2025audio} 
& 77.5 & 0.7 & 0.4 & 0.7 
& 84.4 & 0.0 & 3.0 & 1.6 
& 58.1 & 0.5 & 0.0 & 0.5 
& \textbf{79.3} & 48.5 & 42.8 & 9.8 
\\
Audio-Reasoner~\cite{xie2025audio} 
& 84.7 & 4.6 & 19.0 & 42.0 
& 43.9 & 7.0 & 19.0 & 65.8
& 41.4 & 2.0 & 17.4 & 42.3
& 46.2 & 24.7 & 29.9 & 46.8
\\
Phi-4-Multimodal~\cite{abouelenin2025phi} 
& 24.7 & 2.2 & 76.4 & 78.7 
& 39.4 & 37.3 & 86.6 & 85.9 
& 22.5 & 34.6 & 81.5 & 80.2 
& 62.8 & 56.0 & 82.3 & 88.5 
\\
BALSa-MA~\cite{kuan2025alignment} 
& \textbf{96.4} & \underline{74.9} & 90.2 & 79.5 
& 74.7 & 28.6 & \underline{97.4} & 84.0 
& 40.6 & 39.7 & 85.6 & 80.8 
& 64.6 & \underline{70.7} & 55.8 & 89.8 
\\
\midrule
\textbf{Reasoning Models} 
\\
Qwen2.5-Omni~\cite{xu2025qwen2}
& \textbf{96.4} & 59.1 & \underline{91.6} & \underline{89.6}
& \textbf{92.2} & 51.8 & 94.0 & 91.6
& \textbf{83.1} & \textbf{76.6} & \textbf{90.6} & \textbf{95.3}
& \underline{75.4} & 59.3 & 80.7 & \underline{94.7}
\\
Audio Flamingo 3~\cite{goel2025audio}
& 77.5 & 31.9 & 35.8 & 35.1
& 84.4 & \textbf{69.1} & 72.4 & 39.5
& 58.1 & 60.3 & 56.9 & 47.4 
& \textbf{79.3} & 68.6 & 63.3 & 66.3
\\
BALSa-MA~\cite{kuan2025alignment} 
& \textbf{96.4} & \textbf{96.9} & \textbf{97.8} & 88.9
& 74.7 & \underline{60.4} & \textbf{97.7} & \underline{92.9} 
& 40.6 & 64.2 & \underline{90.2} & 89.0 
& 64.6 & \textbf{74.5} & \textbf{95.1} & 91.8
\\
\midrule
\textbf{Proprietary Models} 
\\
Gemini-2.5-Flash~\cite{comanici2025gemini} 
& 83.6 & 40.2 & 86.4 & 76.1 
& 77.8 & 25.0 & 87.5 & 72.9 
& 58.1 & 33.0 & 86.6 & 76.1 
& 67.6 & \underline{64.0} & \underline{83.1} & 76.9 
\\
Gemini-2.5-Pro~\cite{comanici2025gemini} 
& 93.8 & 55.3 & 82.0 & 64.6 
& 72.3 & 16.2 & 71.8 & 71.7 
& 65.9 & 37.5 & 77.1 & 70.8 
& 71.4 & 54.8 & 50.0 & 58.3 
\\
GPT-4o-Audio~\cite{achiam2023gpt} 
& 86.1 & 51.0 & 86.1 & 80.0 
& \underline{88.9} & 29.4 & 87.2 & 83.8 
& 58.6 & 7.6 & 87.2 & \underline{90.1} 
& 69.4 & 37.2 & 71.9 & 91.3
\\
\bottomrule
\end{tabular}
\end{adjustbox}
\label{tab:main-evaluation-results}
\end{table*}
In this work, we adopt a set of well-known, fully open-source, and extensively documented models as baselines.
These include Qwen-Audio-Chat~\cite{chu2023qwen}, Qwen2-Audio-Instruct~\cite{chu2024qwen2}, Qwen2.5-Omni~\cite{xu2025qwen2}, SALMONN~\cite{tang2023salmonn} (7B and 13B variants), LTU~\cite{gong2023listen}, LTU-AS~\cite{gong2023joint}, GAMA~\cite{ghosh2024gama}, Audio Flamingo 2~\cite{ghoshaudio}, Audio Flamingo 3 (AF3)~\cite{goel2025audio}, Audio-Reasoner~\cite{xie2025audio}, Phi4-Multimodal-Instruct~\cite{abouelenin2025phi}, and BALSa-MA~\cite{kuan2025alignment}.
We also compare with proprietary commercial systems, including Gemini-Flash-2.5~\cite{comanici2025gemini}, Gemini-Pro-2.5~\cite{comanici2025gemini}, and GPT-4o-Audio~\cite{achiam2023gpt}.
Notably, some of the baselines, including Qwen2.5-Omni~\cite{xu2025qwen2}, Audio Flamingo 3~\cite{goel2025audio}, and Audio-Reasoner~\cite{xie2025audio}, are designed with strong joint reasoning capabilities over audio and text. 
Most of these ALLMs are tuned for instruction following and are designed to produce free-form responses rather than select from predefined choices~\cite{chu2023qwen, chu2024qwen2, tang2023salmonn, gong2023listen, gong2023joint}.
To make evaluation consistent, we apply a structured process to extract the relevant answers from the generated outputs.
All experiments use greedy decoding with a maximum output length of 1024 tokens. For answer extraction, we rely on carefully designed regular expressions to identify key information from model responses, similar to the approach followed by MMAU~\cite{sakshi2024mmau}.
In addition, we follow the evaluation protocol defined in~\ref{subsection-evaluation-protocol}: accuracy is reported for answerable tasks, and conditional accuracy (CA) is used for AAD, IASD, and IAQD cases.


\section{Result}

\subsection{Performance on Answerable Questions}
Before evaluating how models handle unanswerable questions, we first measure their performance on standard tasks where a correct answer is always available. 
This initial step allows us to establish a clear baseline and confirm their core capabilities in audio understanding.
As presented in Table~\ref{tab:main-evaluation-results} under the \textit{Ori.} columns, our findings show that many leading ALLMs are highly proficient at answering well-posed questions. 
This is particularly evident in the Animal Sound category, where models like Qwen2.5-Omni and BALSa-MA achieved an outstanding accuracy of 96.4\%. 
Similarly, proprietary models such as Gemini-2.5-Pro also demonstrated strong performance, correctly identifying the animal sound in 93.8\% of cases. 
This high level of accuracy extends to other categories, with models consistently performing well on Vocal and Music Instrument sounds.
Even on the more complex MMAU benchmark, which demands deeper reasoning, top-tier models like Qwen2.5-Omni and AF3 showcase solid results, achieving 75.4\% and 79.3\% respectively. 
This confirms that current state-of-the-art models have developed a robust ability to process audio content and connect it to textual queries in standard, solvable scenarios.
In summary, the strong performance on these answerable tasks demonstrates the advanced state of ALLMs. 
However, this success in ideal conditions raises a critical question: can these models recognize when a question cannot or should not be answered? 
The following subsections explore this challenge, revealing a significant gap between their performance on solvable and unsolvable problems.

\subsection{Degradation on Unanswerable Questions}
While leading models demonstrated proficiency on standard tasks, their performance shifted significantly when faced with the unanswerable questions in AQUA-Bench. 
This reveals a notable performance gap between understanding audio content and recognizing the limits of that understanding. 
Even the most advanced models, which excelled on solvable questions, often struggled to identify when a question was unanswerable.
This challenge is clearly illustrated by some of the top-performing models. 
For example, Audio Flamingo 3 saw its accuracy on the Animal Sound AAD task decrease sharply from 77.5\% on the original task to a score of just 0.7\%. 
This indicates that when the correct option was removed, the model tended to select an incorrect one, even when a none of the above option was provided.
Similarly, Qwen2.5-Omni, a powerful multimodal model, scored an impressive 92.2\% on the standard Vocal Sound task, but this fell to 7.2\% on the AAD variant.
However, this struggle is not uniformly severe across all models, and some exhibit encouraging robustness. In contrast to the models above, BALSa-MA attained 74.9\% accuracy on the same Animal Sound AAD task.
While there is still room for improvement, this indicates a comparatively stronger ability to recognize missing answers.

Furthermore, some models proved adept at specific types of unanswerable questions. 
For instance, both Audio Flamingo 3 and Qwen2.5-Omni scored exceptionally well on the MMAU IAQD task (99.1\% and 90.8\%, respectively), correctly recognizing when a question was fundamentally irrelevant to the audio provided.
This contrast is central to our findings. 
It shows that while a general weakness exists, the ability to handle unanswerable questions varies greatly depending on the model's design and the nature of the challenge. 
A model can be brittle in one scenario (like a missing answer choice) yet robust in another (like an irrelevant question). 
This suggests that models are not failing uniformly but have specific, distinct blind spots.
To understand these nuances, the following section will delve into the three unanswerable scenarios, analyzing why models succeed in some areas while failing dramatically in others.

\subsection{Analysis of Unanswerable Scenarios}
To understand the causes of the performance decline, we analyze how the models behave within each of the three distinct unanswerable scenarios. 
This reveals that models do not fail uniformly; rather, each scenario exposes a different type of weakness in their design and training.
The Absent Answer Detection task is arguably the most straightforward challenge: the model is presented with a clear audio clip and a valid question, but the correct answer is simply missing from the options. 
The ideal response is to recognize this absence.
However, our results in Table~\ref{tab:main-evaluation-results} show that this is where most models struggle the most. 
We observe a strong tendency toward a ``forced-choice,'' with models often selecting one of the provided options rather than choosing the none of the above alternative.
For example, Qwen2.5-Omni, which achieved 96.4\% on the original Animal Sound task, dropped to 20.5\% on the AAD version.
This suggests that the model's internal confidence in an incorrect answer was higher than its confidence that no correct answer was present. 
This behavior was even more pronounced in models like Audio Flamingo 3, which fell from 77.5\% to 0.7\% on the same task.

In the Incompatible Answer Set Detection scenario, the mismatch is more abstract. 
The answer options belong to a semantic category that is completely irrelevant to the question. 
This tests a model's ability to grasp categorical relationships.
Interestingly, models performed significantly better on this task compared to AAD. 
For instance, Qwen2.5-Omni achieved high scores across multiple IASD tasks, including 83.6\% for Animal Sounds and 87.1\% for Vocal Sounds. 
Likewise, BALSa-MA scored an exceptional 97.4\% on the Vocal Sound IASD task. 
This widespread success suggests that modern ALLMs have a strong internal understanding of semantic categories. 
They can effectively recognize that ``happy'' is not a valid answer for the question ``What animal is making this sound?''. 
This finding shows their knowledge is not just superficial pattern matching but is organized around meaningful concepts.

The Incompatible Audio Question Detection task presents the most complex challenge, requiring models to reason about what information can and cannot be inferred from audio alone. 
For example, a model must know that the color of a dog cannot be determined from the sound of its bark.
Performance on this task clearly separates the models based on their reasoning capabilities. 
We see a distinct group of high-performers that seem to possess this deeper, cross-modal understanding. Audio Flamingo 2 and Qwen2.5-Omni delivered outstanding results on the MMAU IAQD task, with scores of 99.1\% and 90.8\%, respectively. 
This implies they have robust world knowledge and can identify when a question is unanswerable due to the limitations of the audio modality.
In contrast, other models struggled, likely resorting to guessing or hallucinating an answer based on common associations. 
The polarized results on the IAQD task highlight its effectiveness as a probe for the advanced reasoning abilities that are essential for creating truly trustworthy AI systems.

To further probe these behaviors, we also explored the effect of including explicit guidance in the prompt. 
For instance, we appended instructions such as, ``Select None of the above if you believe none of the listed answers are right'' for AAD and IASD tasks, or Pick ``Unanswerable when the audio lacks the details needed to decide'' for IAQD tasks. We observed an universal improvement in performance across all models. 
This finding suggests that the models possess the latent ability to identify unanswerable questions, but often fail to apply this capability without direct instruction. 
Their default behavior exhibits a forced-choice bias, tending to select an option even when none is appropriate.
Due to space constraints, we present these results on our demo website~\footref{demo}.

\subsection{Reasoning Models on Unanswerables}
A key question is whether models renowned for their advanced reasoning capabilities, such as Audio Flamingo 3 (AF3) and Qwen2.5-Omni, are inherently more robust against unanswerable questions. 
To investigate this, we evaluated these models, along with BALSa-MA, using a Chain-of-Thought (CoT)~\cite{wei2022chain} prompting strategy. 
We instructed the model to first reason step-by-step about whether the question is answerable based on the provided audio.
The results in Table~\ref{tab:main-evaluation-results} reveal substantial improvements across all three systems. AF3, which initially scored near-zero on several AAD tasks, recovered strongly with CoT, reaching 31.9\% on Animal Sounds and 69.1\% on Vocal Sounds, and achieving 68.6\% on MMAU. 
Qwen2.5-Omni also showed large gains, with its Animal Sound AAD score rising from 20.5\% to 59.1\% and its overall MMAU AAD accuracy improving from 28.3\% to 59.3\%. 
BALSa-MA, already strong without explicit reasoning, further advanced to stronger performance under CoT, consistently exceeding 90\% on most unanswerable settings.
These findings convey two insights. 
First, strong reasoning models indeed possess a latent ability to identify unanswerable questions, but tend to default to forced-choice behavior unless explicitly guided. 
Second, while CoT helps all models, its effect is particularly striking on AF3 and Qwen2.5-Omni, which transition from near-random guessing to competitive accuracy. 

\section{Conclusion}
We present AQUA-Bench, a benchmark for evaluating unanswerability in audio question answering through three scenarios: Absent Answer Detection, Incompatible Answer Set Detection, and Incompatible Audio Question Detection. 
Experiments show that while ALLMs excel on standard answerable tasks, they suffer from a pronounced forced-choice bias, often answering when they should abstain. 
We further demonstrate that explicit prompting and reasoning can recover much of this hidden ability. 
AQUA-Bench thus exposes a critical blind spot and provides a principled tool for building audio-language systems that are both accurate and trustworthy.

\section{Acknowledgment}
This work was supported by the Ministry of Education (MOE) of Taiwan under the project Taiwan Centers of Excellence in Artificial Intelligence, through the NTU Artificial Intelligence Center of Research Excellence (NTU AI-CoRE).
In addition, we thank the National Center for High-performance Computing (NCHC) of the National Applied Research Laboratories (NARLabs) in Taiwan for providing computational and storage resources.

\section{Discussion on Methodology and Limitations}
\label{sec:appendix_discussion}

In this section, we provide further context regarding our experimental design choices, data quality control, and current limitations.

\subsection{Rationale for Experimental Design}
Our benchmark is designed with specific constraints to strictly isolate and evaluate a model's ability to handle unanswerable queries. 
\newline

\noindent\textbf{Simplicity of Underlying Tasks:} We intentionally constructed unanswerable variants on top of standard, widely used audio understanding tasks. The simplicity of these tasks is a deliberate design choice. It allows us to confirm that a model's performance drop is caused by the introduced unanswerability, rather than a lack of fundamental audio perception skills. \newline

\noindent\textbf{Multiple-Choice Format:} We utilized a multiple-choice framework to establish a controlled environment. This format provides a standardized metric to precisely measure whether a model can correctly abstain from answering. While free-form generation is more natural, it often introduces ambiguity in evaluation. The multiple-choice setting ensures a clear signal for measuring ``unanswerability'' without the need for complex judge models.

\subsection{Data Quality and Qualitative Analysis}
We acknowledge the importance of rigorous data validation and qualitative insights. To address concerns regarding reliance on GPT-4o and the need for qualitative examples, we implemented the following measures:
\newline

\noindent\textbf{Mitigating Model Bias:} To reduce the influence of GPT-4o's stylistic patterns, we manually annotated a seed set of examples. These were used as in-context demonstrations to guide the data generation process, ensuring more diverse and accurate outputs.
\newline
    
\noindent\textbf{Human Verification:} We conducted human verification on a subset of the generated data. 
\newline
    
\noindent\textbf{Qualitative Examples:} We agree that qualitative analysis is crucial for understanding model behavior. We have hosted a set of examples on our project website~\footnote{\scriptsize\url{https://github.com/kuan2jiu99/aqua-bench}}. 
These examples illustrate typical failure modes, such as hallucinations, across different unanswerable scenarios.

\subsection{Limitations and Future Directions}
While our current framework effectively highlights the problem of hallucination under unanswerability, we identify two key areas for future improvement:
\newline

\noindent\textbf{Scenario Complexity:} Our current questions focus on controlled audio-question mismatches. Real-world scenarios are often more complex, involving queries about visual attributes, specific speaker identities, or object appearances that cannot be inferred from audio alone. Extending the evaluation to cover these ``in-the-wild'' scenarios is a priority for our future work.
\newline
    
\noindent\textbf{Open-Ended Generation:} Although the multiple-choice format is effective for diagnostic purposes, we recognize that open-ended generation better reflects real-world user interactions. Future research will explore robust evaluation protocols for free-form responses, potentially utilizing advanced LLMs as judges to assess how well models abstain in open-ended settings.


\bibliographystyle{IEEEbib}
\bibliography{strings,refs}

\end{document}